\documentclass[12pt]{article}
\usepackage{graphicx}
\usepackage{times}
\usepackage{setspace}
\usepackage{authblk}
\usepackage{pseudocode} 
\begin{document}
\title{Towards Exascale Lattice Boltzmann computing} 
\author[1,2,5]{S. Succi}
\author[3]{G. Amati}
\author[5]{M. Bernaschi} 
\author[4,2]{G. Falcucci}
\author[5]{M. Lauricella}
\author[5]{A. Montessori}
\affil[1]{Italian Institute of Technology, Center for Life Nanoscience at La Sapienza,  00161, Rome, Italy}
\affil[2]{John A. Paulson School of Engineering and Applied Sciences, Harvard University, 33 Oxford St., 02138 Cambridge, MA, USA}
\affil[3]{SCAI, SuperComputing Applications and Innovation Department, CINECA, Via dei Tizii, 6, 00185 Rome, Italy}
\affil[4]{Department of Enterprise Engineering ``Mario Lucertini'', University of Rome “Tor Vergata”, Via del Politecnico 1, 00133 Rome, Italy}
\affil[5]{Istituto per le Applicazioni del Calcolo CNR, Via dei Taurini 19, 00185 Rome, Italy}
\maketitle

\newpage

\tableofcontents

\newpage

\listoffigures

\newpage

\section*{Abstract}
We discuss the state of art of Lattice Boltzmann (LB) computing, with special
focus on prospective LB schemes capable of meeting the forthcoming Exascale challenge.
After reviewing the basic notions of LB computing, we discuss current techniques to improve the performance
of LB codes on parallel machines and illustrate selected leading-edge applications in the Petascale range.
Finally, we put forward a few ideas on how to improve the  communication/computation overlap in current 
large-scale LB simulations, as well as possible strategies towards fault-tolerant LB schemes.    

\section{Introduction}
Exascale computing refers to computing systems capable of delivering Exaflops, one billion billions ($10^{18}$, also known as {\it quintillion}) floating-point operations per second, that is one floating-point operation every billionth of a billionth of second, also known as Attosecond ($10^{-18}$ s). 
Just to convey the idea, at Attosecond resolution, one can take a thousand snapshots of 
an electron migrating from one atom to another to establish a new chemical bond.
Interestingly, the attosecond is also the frontier of current day atomic clock precision, which means clocks that lose or gain less than two seconds over the entire age of the Universe!\\ 
In 2009 Exascale was projected to occur in 2018, a prediction which turned out to be fulfilled just months ago,
with the announcement of a 999 PF/s sustained throughput computational using the SUMMIT supercomputer~\cite{ref:SUMMIT}\footnote{This figure was obtained using half precision (FP16) floating point arithmetic.}.   
Mind-boggling as they are, what do these numbers imply for the actual advancement of science?
The typical list includes new fancy materials, direct simulation of biological organs, fully-digital design of cars and airplanes with no need of building physical prototypes, a major boost in cancer research, to name but a few.   \\
Attaining Exaflop performance on each of these applications, however, is still an open challenge,
 because it requires a virtually perfect match between the system architecture and
the algorithmic formulation of the mathematical problem.\\
In this article, we shall address the above issues with specific reference to a class of mathematical
models known as Lattice Boltzmann (LB) methods, namely a lattice formulation of Boltzmann's kinetic equation
which has found widespread use across a broad range of problems involving complex states of flowing
matter \cite{ref:LB1}.\\
The article is organized as follows: in Section~\ref{LBE_intro} we briefly introduce the basic features of the LB method.
In Section~\ref{improvements} the state of-the-art of LB performance implementation will be briefly presented, 
whereas in Section ~\ref{Exascale} the main aspects of future exascale computing are discussed,  
along with performance figures for current LB schemes on present-day Petascale machines.
Finally, in Section~\ref{Future} we sketch out a few \textit{tentative} ideas which may facilitate the migration of LB
codes to Exascale platforms.

\section{Basic Lattice Boltzmann Method} \label{LBE_intro}
\begin{figure}[h]
\includegraphics[scale=0.4]{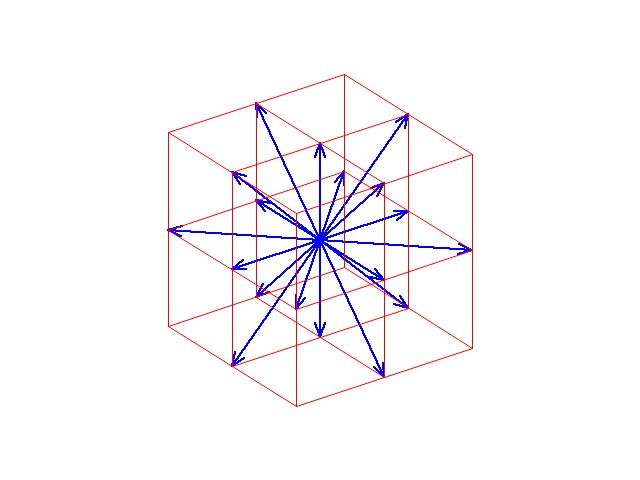} 
\includegraphics[scale=0.4]{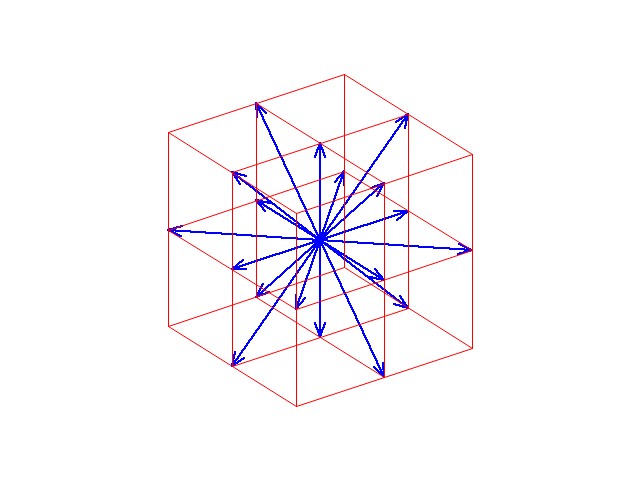} 
\caption{D3Q27 lattice, composed of a set of 27 discrete velocities (left). 
         D3Q19 lattice, composed of a set of 19 discrete velocities (right). }  \label{fig:D3Q27}
\end{figure}

The LB method was developed in the late 1980's as a noise-free replacement of lattice gas cellular automata 
for fluid dynamic simulations \cite{ref:LB1}.\\
Ever since, it has featured a (literal) exponential growth of applications across
a remarkably broad spectrum of complex flow problems, from fully developed turbulence
to micro and nanofluidics \cite{ref:AM2015}, all the way down to quark-gluon plasmas \cite{ref:LB2008,ref:LB2038}.\\
The main idea is to solve a minimal Boltzmann kinetic equation for a set
of discrete distribution functions (\textit{populations} in LB jargon) $f_i(\vec{x};t)$, expressing
the probability of finding a particle at position $\vec{x}$ and time $t$, with a (discrete) velocity  $\vec{v}=\vec{c}_i$. 
The set of discrete velocities must be chosen in such a way as to secure enough symmetry 
to comply with mass-momentum-energy conservation laws of macroscopic hydrodynamics, as well as
with rotational symmetry.
Figure \ref{fig:D3Q27} shows two of the 3D lattices most widely used for current LB simulations, with a set of 19 discrete velocities (D3Q19) or 27 velocities (D3Q27). \\
In its simplest and most compact form, the LB equation reads as follows:
\begin{equation} \label{eq:eq1}
f(\vec{x}+\vec{c}_i,t+1) = f'_i(\vec{x};t) \equiv (1-\omega) f_i(\vec{x};t) + \omega f_i^{eq}(\vec{x};t) + S_i,\;\;i=0,b
\end{equation}
where $\vec{x}$ and $\vec{c}_i$ are 3D vectors in ordinary space, $ f_i^{eq} $ is the equilibrium distribution function 
and the lattice time step is made unit, so that $\vec{c}_i$ is the length of the link connecting a generic lattice site
node $\vec{x}$ to its $b$ neighbors, $\vec{x}_i = \vec{x}+\vec{c}_i$.\\
In the above, the local equilibria are provided by a lattice truncation, to
second order in the Mach number $M=u/c_s$, of the Maxwell-Boltzmann distribution, namely
\begin{equation} \label{eq:equil}
f_i^{eq}(\vec{x};t) = w_i \rho (1 + u_i + q_i)
\end{equation}
where $w_i$ is a set of weights normalized to unity, 
$u_i = \frac{\vec{u} \cdot \vec{c}_i}{c_s^2}$ and $q_i = u_i -u^2/D$, in $D$ spatial dimensions.\\
Finally, $S_i$ is a source term encoding the fluid interaction with external (or internal) sources, the latter
being crucial to describe multiphase flows with potential energy interactions.\\
The above equation represents the following situation: the populations at site $\vec{x}$ at time $t$ 
collide to produce a post-collisional state $f'_i(\vec{x};t)$, which is then scattered away 
to the corresponding neighbors at $\vec{x}_i$ at time $t+1$.
Provided the lattice is endowed with suitable symmetries, and the local equilibria
are chosen accordingly, the above scheme can be shown to reproduce the 
Navier-Stokes equations for an isothermal quasi-incompressible fluid of density and velocity
\begin{equation} \label{eq:eq2}
\rho = \sum_i f_i  \hspace{2cm}
\vec{u}=(\sum_i f_i \vec{c}_i)/\rho
\end{equation}
The relaxation parameter $\omega$ dictates the viscosity of the lattice fluid according to
\begin{equation} \label{eq:eq3}
\nu = c_s^2 (\omega^{-1}-1/2)
\end{equation}
Full details can be found in the vast literature on the subject~\cite{ref:LB2.1}\cite{ref:LB2.2}.

\subsection{The stream-collide LB paradigm}
The main strength of the LB scheme is the stream-collide paradigm, which stands
in marked contrast with the advection-diffusion of the macroscopic 
representation of fluid flows.\\
Unlike advection, streaming proceeds along straight lines defined by the discrete
velocities $\vec{c}_i$, regardless of the complexity of the fluid flow. 
This is a major advantage over material transport
taking place along fluid lines, whose tangent is given by the fluid velocity itself, a highly
dynamic and heterogeneous field in most complex flows.
To be noted that streaming is literally exact. i.e., there is no round-off error, as it
implies a memory shift only with no floating-point operation.\\
For the case of a 2D lattice, discounting boundary conditions, the streaming along positive 
and negative $v_x$ (p=1, 2) can be coded as shown in Algorithms ~\ref{pseudo1} and ~\ref{pseudo2}.\\
Note that the second loop runs against the flow, i.e. counterstream, so as to avoid overwriting 
memory locations, which would result in advancing the same population across the 
entire lattice in a single time step!
A similar trick applies to all other populations: the $(i,j,k)$ loops
must run counter the direction of propagation.\\
This choice makes possible to work with just a single array for time levels $t$ and $t+1$, thereby halving the memory requirements.\\
With this streaming in place, one computes the local equilibria distribution $f_i^{eq} $ as a 
quadratic combination of the populations and moves on to the collision step, as shown in pseudocode \ref{pseudo3}. 
This completes the inner computational engine of LB.
In this paper, we shall not deal with boundary conditions, even though, like with any
computational fluid-dynamic method,  they ultimately decide the quality of the simulation results.
We highlight only that the exact streaming permits to handle fairly complex
boundary conditions in a {\it conceptually} transparent way~\cite{ref:BC} because information always moves along the straight lines defined by the discrete velocities.
Actual programming can be laborious, though.\\ 
\----------------------------------------------------------------------------------------------\\
\begin{pseudocode}{In-place streaming $ v_x > 0 $}{.} \label{pseudo1}
\FOR j \GETS 1 \TO ny \\
\FOR i \GETS 1 \TO nx \\ 
     f(1,i,j) = f(1,i-1,j)
\end{pseudocode}
\\
\----------------------------------------------------------------------------------------------\\
\begin{pseudocode}{In-place streaming $ v_x < 0 $}{.} \label{pseudo2}
\FOR j \GETS 1 \TO ny \\
\FOR i \GETS nx \TO 1 \\ 
     f(2,i,j) = f(2,i+1,j)
\end{pseudocode}
\\
\----------------------------------------------------------------------------------------------\\
\begin{pseudocode}{Collision step}{.} \label{pseudo3}
\FOR j \GETS 1 \TO ny \\
\FOR i \GETS 1 \TO nx \\
\FOR p \GETS 1 \TO npop \\
     f(p,i,j)=(1-\omega)*f(p,i,j,k)+\omega*f^{eq}(p,i,j)
\end{pseudocode}
\section{Improving LB performance} \label{improvements}
The LB literature provides several ``tricks'' to accelerate the execution of the basic LB algorithms above described.
Here we shall briefly describe three major ones, related to optimal data storage and
access, as well as parallel performance optimisation.

\subsection{Data storage and allocation}
The Stream-Collide paradigm is very powerful, but due to the comparatively large number
of discrete velocities, it shows exposure to memory bandwidth issues, i.e. the number 
of bytes to be moved to advance the state of each lattice site from time $t$ to $t+1$.
Such data access issues have always been high on the agenda of efficient LB implementations,
but with prospective Exascale LB computing in mind, they become vital.\\
Thus, one must focus on efficient data allocation and access practices.
There are two main and mutually conflicting ways of storing LB populations (assuming row-major memory order as in
C or C++ languages):
\begin{itemize}
\item By contiguously storing the set of populations living in the same lattice site: \verb|fpop[nx][ny][npop]|(AoS, Array of Structures) 
\item By contiguously storing homologue streamers across the lattice:\\
\verb|fpop[npop][nx][ny]| (SoA, Structure of Arrays)
\end{itemize}
The former scenario, AoS, is optimal for collisions since the collision
step requires no information from neighbouring populations and it is also 
cache-friendly,  as it presents unit stride access, i.e., memory 
locations are accessed contiguously, with no jumps in-between.\\
However, AoS is non-optimal for streaming, whose stride is now equal 
to \verb|npop|, the number of populations.\\
The latter scenario, SoA, is just reciprocal, i.e., optimal for streaming, since it features a unitary stride, but 
inefficient for collisions, because it requires \verb|npop| different streams of data at the same time.\\
These issues are only going to get worse as more demanding applications
are faced, such multiphase or multicomponent flows requiring more populations
per lattice site and often also denser stencils. 
Whence the need of developing new strategies for optimal data storage and access. 
A reliable performance estimate is not simple, as it depends on a number of hardware platform parameters, primarily
the latency and bandwidth of the different levels of memory hierarchy.\\
An interesting strategy has been presented by Aniruddha and coworkers \cite{ref:ANSUMALI},
who proposed a compromising data organization, called AoSoA.
Instead of storing all populations on site, they propose to store only relevant
subsets, namely the populations sharing the same energy (energy-shell organization).
For instance, for the D3Q27 lattice, this would amount to four energy shells: 
energy $0$ (just 1 population), energy $1$ (6 populations), 
energy $2$ (12 populations)  and energy $3$ (8 populations).
This is sub-optimal for collisions but relieves much strain from streaming.
As a matter of fact, by further ingenuous sub-partitioning of the energy shells, the authors manage to capture the best of the two, resulting in an execution of both Collision and Streaming
at their quasi-optimal performance. \\
Note that in large-scale applications, streaming can take up to $30$ per cent of the execution time
even though it does not involve any floating-point operation!\\
In ~\cite{ref:Array} a thorough analysis of various data allocation policies is presented. 

\subsection{Fusing Stream-Collide}
Memory access may take a significant toll if stream and collide are kept as separate steps (routines). 
Indeed, node-wise, using a single-phase double precision D3Q19 LB scheme, there are 
$19 \times 8 \times 2 = 304$ bytes involved in memory operations (load/store) per $ \simeq 250 $ 
Flops of computation, corresponding to a rough computation to memory ratio of $\simeq 0.7$ Flops/Byte.\\
The memory access issue was recognized since the early days of 
LB implementations and led to the so-called fused, or push, scheme.
In the fused scheme (eq.~\ref{eq:eq4}), streaming is, so to say, ``eaten up'' by collisions
and replaced by non-local collisions \cite{ref:Mattila}.
\begin{equation} \label{eq:eq4}
f'(p,i,j,k;t+1) = (1-\omega)*f(p,ip,jp,kp;t)+ \omega*f^{eq}(p,ip,jp,kp;t)
\end{equation} 
where $ (ip,jp,kp)$ is the pointee of site $(i,j,k)$ along the $p$-th direction.
As one can appreciate, the streaming has virtually disappeared, to be
embedded within non-local collisions, at the price of doubling memory requirements (two separate arrays at $t$ and $t+1$ are needed) and a more complex load/store operation.\\
Actually, fused versions with a {\em single} memory level have also appeared in the literature,
but it is not clear whether their benefits are worth the major programming complications they entail \cite{ref:Massa}.
Moreover the chance of executing them in parallel mode is much more limited.

\subsection{Hiding communication costs} \label{Hiding}
The third major item, which is absolutely crucial for multiscale applications, where
LB is coupled to various forms of particle methods, is the capability of hiding the
communication costs by overlapping communication and calculation when running in parallel mode.
Technically, this is achieved by using non-blocking communication 
send/receive MPI primitives as shown in pseudo-algorithm~\ref{pseudo4}:
\\
\----------------------------------------------------------------------------------------------\\
\begin{pseudocode}{non-blocking Receive}{.} \label{pseudo4}
Source Task  (st): Send(B,dt)\\
Source Task (st): Do\_some\_work() \\
Destination Task  (dt): Receive(B,st) 
\end{pseudocode}
\\
where $B$ is the boundary data to be transferred from source to destination, in order 
for the latter to perform its chunk of computation.
The send operation is usually non-blocking, i.e. the source task keeps doing other 
work while the memory sub-system is transmitting the data, because these data are, so to say, not its concern.
The destination task, on the other hand, cannot start the calculation until
those data are received, hence the safe procedure is to sit idle 
until those data are actually received, i.e. the blocking-receive scenario.\\
However, if other work, not contingent to the reception of those data is 
available, there is obviously no point of waiting, whence the point of non-blocking receive.        
In other words, after posting its request for data from another task, the given
task proceeds with the calculations which are free from causal dependencies on the
requested data, until it is notified that these data are finally available.\\
This is of course way more complicated than just sitting idle until the requested data
arrive, as it requires a detailed command at the concurrent timeline of the parallel code.
However, non-blocking receive operations are absolutely key: without hiding communication costs, multiscale LB
applications would not get any near to the extreme performance they have attained in the last decade.
As an example, the multiscale code MUPHY~\cite{ref:Muphy} 
scaled from tens of Teraflops in 2011 to 20 Petaflops in 2013, providing a precious 
asset for the design of future exascale LB code.\\
A description of the state of the art of HPC LB can be found in \cite{ref:WALBEerla}.

\section{Petascale and the roadmap to Exascale computers} \label{Exascale}
Back in 1975, Gordon Moore predicted doubling of computing power every 18 months, namely
a thousandfold increase every fifteen years~\cite{ref:Moore}. 
A unique case in the history of technology, such
prediction, now going by the name of Moore's law, has proved correct 
for the last four decades, taking from CRAY's Megaflops of the early 80's 
to the current two hundreds Petaflops of Summit supercomputer \cite{ref:top500}.\\ 
Besides decreasing single-processor clock time, crucial to this achievement 
has proven the ability of supercomputer architectures to support concurrent 
execution of multiple tasks in parallel.\\
The current leading entries in the list of top-500 performers support up to
millions of such concurrent tasks and each CPU hosts tens of basic computing units, known as ``cores''.
The promised 2020's Exascale machine will cater to hundreds of 
million of cores, with a core-clock still in the GHz range, because of heat power constraints.
As a consequence, an Exascale code should be prepared to exploit up to {\it a billion} concurrent 
cores, each of them able to perform billions of floating point operations: $10^{18} = 10^9 \times 10^9$.
This ``symmetry'' between computation and number of units that need to communicate
reflects and mandates a perfect balance between the two.\\
Indeed, as performance ramps up, it is increasingly apparent that accessing 
data could become more expensive than make computations using them: 
{\it Bytes/s} is taking over {\it Flops/s} as a metrics for extreme performance.\\
Besides hardware issues, Exascale computing also bears far-reaching implications on computational science, setting a strong premium on algorithms
with low data traffic, low I/O, lean communication to computation ratios and high fault-tolerance.
The point is simple, but worth a few additional comments.\\
Let $F$ being the number of Flops required by a given application, $B$ the corresponding number of Bytes, 
$\dot F$ the processing speed (Flops/s) and $\dot B$ (Bytes/s) the corresponding bandwidth. 
On the assumption of no overlap between memory access and computation, the execution time 
is given by the sum of the two, namely:
\begin{equation} \label{eq:eq5}
t = \frac{F}{\dot F} + \frac{B}{\dot B} 
\end{equation}
The corresponding effective processing rate is given by:
\begin{equation} \label{eq:eq6}
\dot F_{eff} \equiv F/t = \frac{\dot F}{1+ \phi}  
\end{equation}
where
\begin{equation} \label{ROOF}
\phi \equiv \frac{\dot F/\dot B}{F/B} 
\end{equation}
This shows that in order to attain the peak performance $\dot F$, corresponding to
no data access overheads (infinite bandwidth), one must secure the condition $\phi \ll 1$. 
Indeed, for $\phi \ll 1$, i.e., a ``large'' $F/B$,  we have $\dot F_{eff}\sim \dot F$, whereas in the opposite limit
$\phi \gg 1$, a ``small'' $F/B$, the effective processing rate flattens at $\dot F_{eff} \sim (F/B) \;\dot B $, i.e., 
a fraction of the bandwidth. 
This is a (simplified) version of the so-called roofline model \cite{ref:ROOFLINE}.\\
In a symbolic form, the roofline condition (\ref{ROOF}) can be written as
 ``{\it Flops/Bytes much larger than (Flops/s)/(Bytes/s)}''. 
The former depends solely on the application, whereas the
latter is dictated by the hardware instead. \\
The ``threshold'' value, at which the transition between memory-bound to 
floating-point-bound regimes occurs, is, for present hardware (fig.~\ref{fig:roof}), $F/B \sim 10$.

\subsection{Vanilla LB: back of envelope calculations} \label{VLB}
\begin{figure}[h] \label{fig:roof}
\includegraphics[scale=0.8]{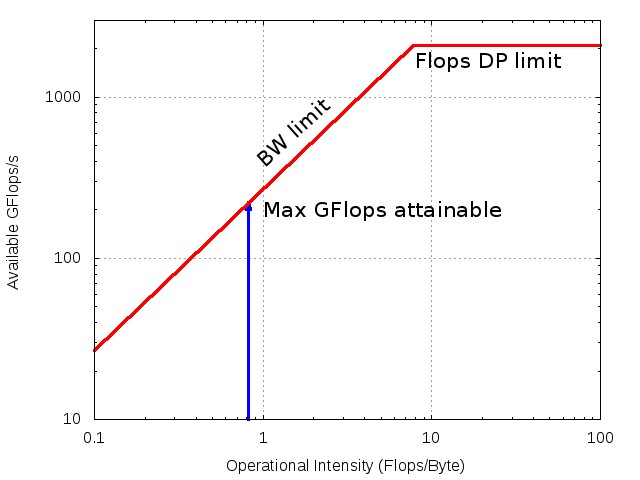}
\caption{Roofline Model obtained for an intel Phi processor: for $F/B<10$ the effective processing speed grows linearly with $F/B$, whereas for $F/B>10$ it comes to a saturation, dictated by the raw processing rate.
Increasing $F/B$ much beyond, does not win any additional performance. The vertical line indicates the performance range for a "Vanilla Lattice Boltzmann".}
\end{figure}

Using a simple {\it Vanilla} LB (VLB), i.e. single-phase, single time relaxation, 3D lattice with fused collision,  
the roofline model permits to estimate limiting figures of performance. In Fig.~\ref{fig:roof} the roofline for an Intel phi is shown\footnote{Bandwidth and Float point computation limits are obtained performing {\em stream} e HPL benchmark}.\\
The technical name of the ratio between flops and data that need to be loaded/stored from/to memory is
Operational Intensity, OI for short, but we prefer to stick to our $F/B$ notation to keep 
communication costs in explicit sight.\\
It provides information about code limitations: whether it is bandwidth-bound 
($F/B < 10$) or floating-point bound ($ F/B > 10$), as shown in Fig.~\ref{fig:roof}.
The HPL code used to rank supercomputers~\cite{ref:top500} is floating-point bound 
\footnote{All hardware is built to exploit performance for floating point bound codes.}.\\ 
For a VLB, $F  \simeq 200 \div 250 $ is the number of floating point operations per lattice site and time step
and $B= 19*2*8=304 $ is load/store demand in bytes, using double precision, hence
$F/B \sim 0.7$, which spells ``bandwidth-bound''.   
Boundary conditions operations are not considered here, for the sake of simplicity.\\
Memory Bandwidth in current-day hardware, at single node level, is about $\simeq 400$ GB/s for an Intel Xeon Phi 7250 using MCDRAM in flat mode. MCDRAM is faster than RAM memory but it is limited in size ($16$ GB). It can be used as a further cache level (cache mode) or auxiliary memory (flat mode)~\cite{ref:IntelKNL}.\\ 
For a ``standard server'' like Intel Xeon Platinum 8160 server Memory Bandwidth is $ \simeq 150 $ GB/s.\\ 
According to technological trends, bandwidth is not expected to exceed 500 GB/s by 2020 \cite{ref:Stream}.
Using as LB performance metrics GLUPS (Giga Lattice Update Per Second), due to the fact that LB is memory bounded the peak value is 400 GB/s, the limit is $ 400 (GB/s) /  305 (B/LUPS) $ yielding $ 1.3 $ GLUPS for a Phi node, so that the
ratio $\dot F /\dot B \sim 5$, yielding an overhead factor $\phi \sim 5/0.7 \sim 7$: the effective processing speed is $\dot F_{eff} \sim 0.125 \dot F $.\\
For an Exascale machine with order of $10^5$ nodes and $ 500 $ GB/s Bandwidth per node, a peak of $ \sim 2 \times 10^{14}$ LUPS is expected.

\subsection{Present-day Petascale LB application}
\begin{figure}[h] 
\begin{center}
\includegraphics[scale=0.22]{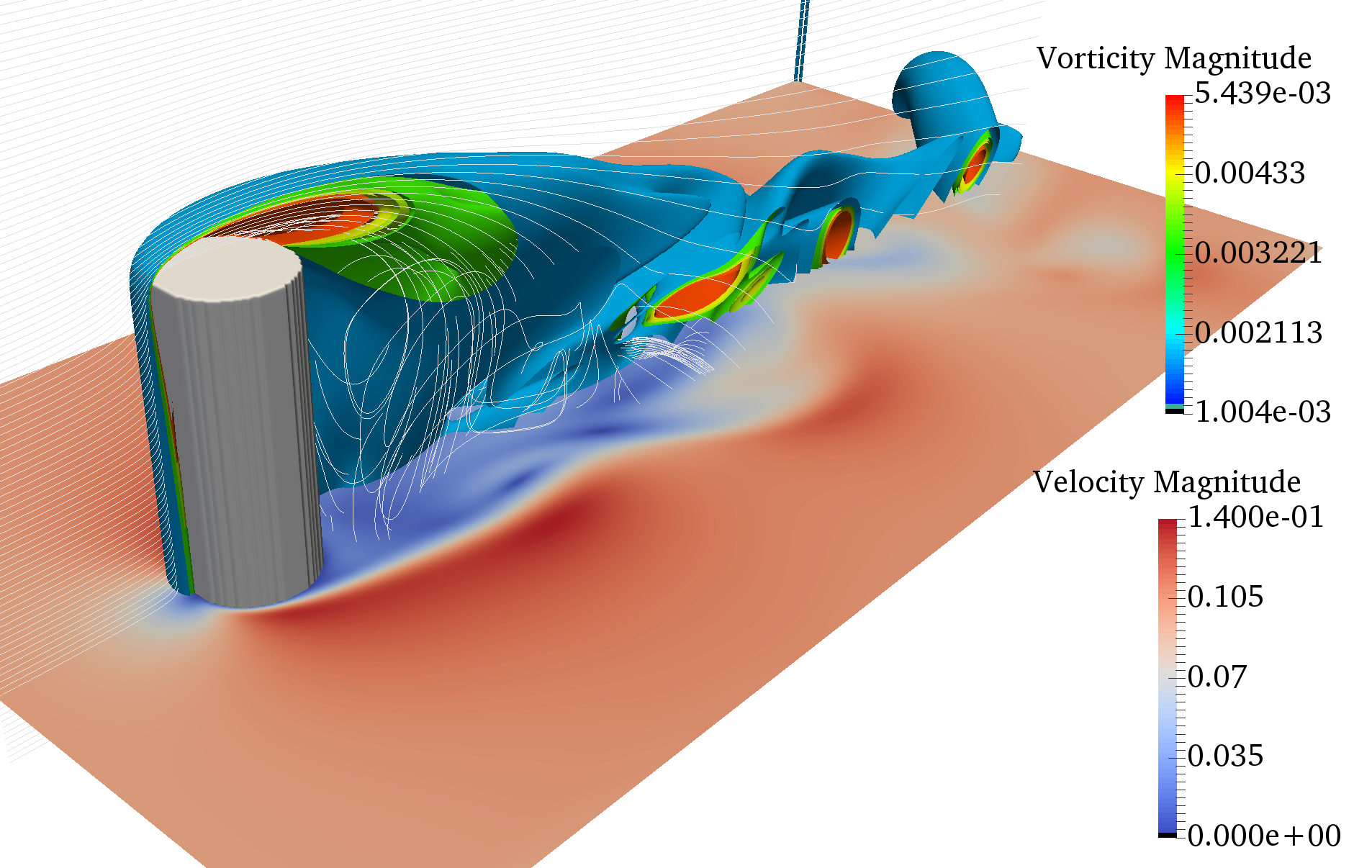}
\caption{Vorticity and velocity contour for the flow around a cylinder}
\label{fig:vorticity2}
\end{center}
\end{figure}

\begin{figure}[h]
\begin{center}
\includegraphics[scale=0.35]{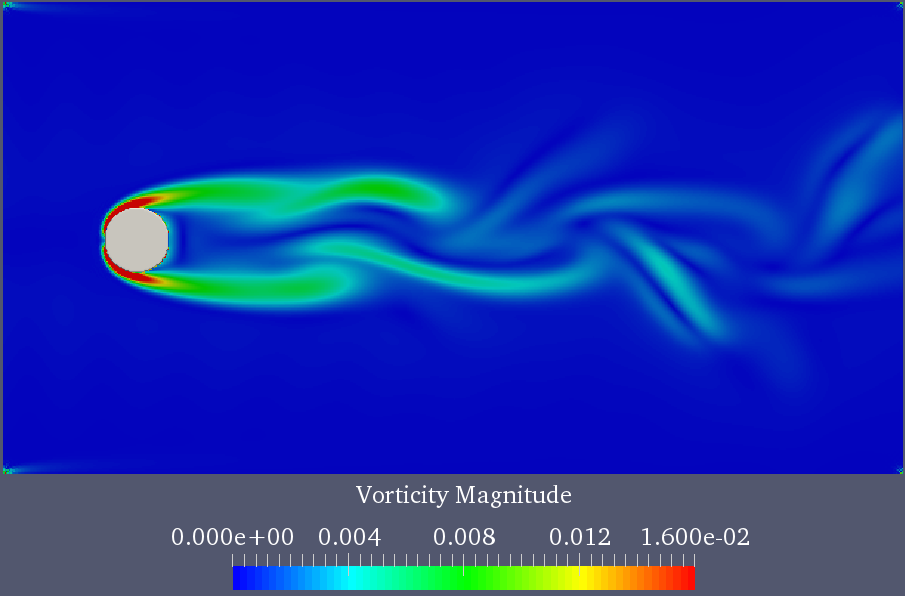}
\includegraphics[scale=0.35]{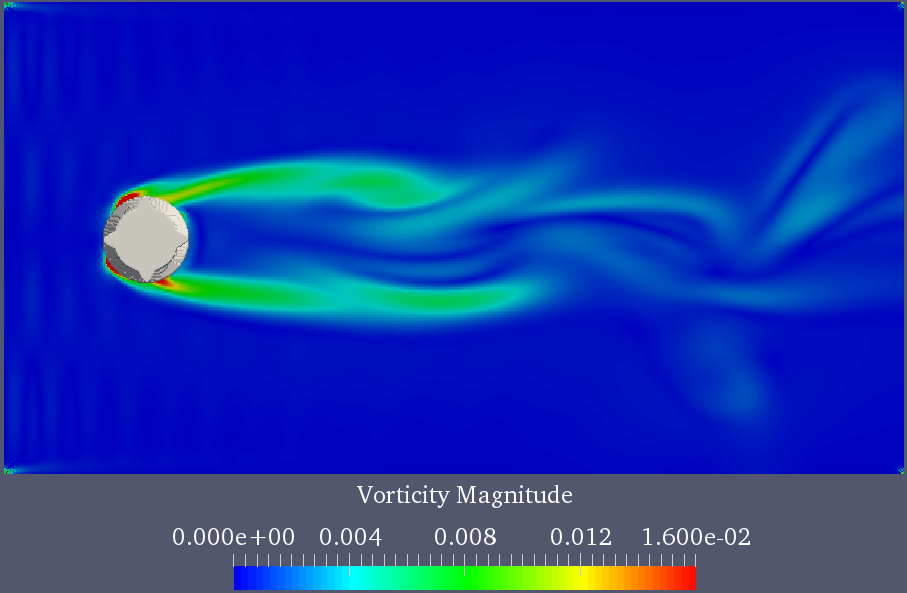}
\caption{Vorticity for a ``clean'' cylinder (top) and a ``decorated'' one (bottom).} \label{fig:vorticity1}
\end{center}
\end{figure}

\begin{figure}[h] \label{fig:speedup1}
\includegraphics[scale=0.8]{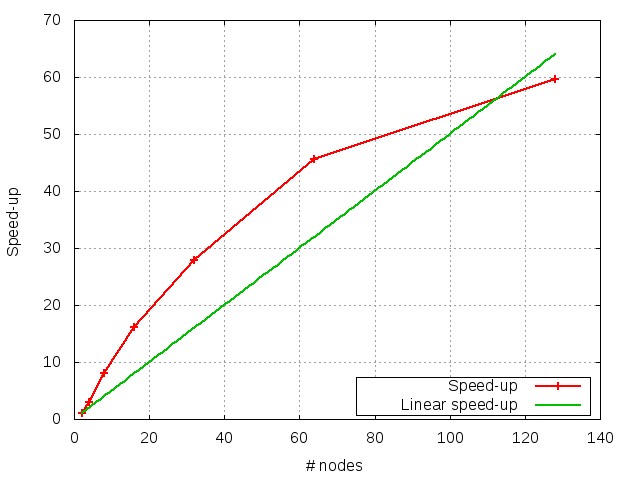}
\caption{Strong scaling for a $ 512^3 $ gridpoint simulation. Note the superlinear speed-up achieved up to 128 nodes (i.e. 512 task).}
\end{figure}

\begin{figure}[h] \label{fig:scaleup3}
\includegraphics[scale=0.8]{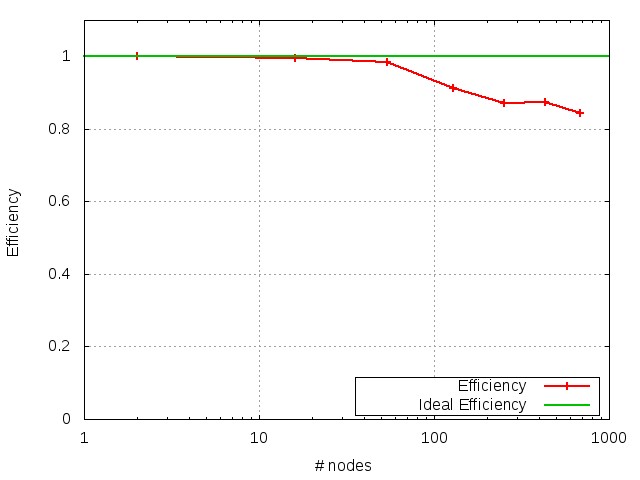}
\caption{Weak scaling using $ 128^3 $ gridpoint per task. Using $2744$ tasks ($ 686 $ nodes) $ 84\% $ of parallel efficiency is obtained.}
\end{figure}

As a concrete example of current-day competitive LB performance, we discuss the case of the
flow around a cylinder (Fig.~\ref{fig:vorticity2},~\ref{fig:vorticity1}): the goal is to study the effect on vorticity 
reduction using {\it decorated} cylinders~\cite{ref:SPONGES} . 
To describe short-scale decorations, such as those found in nature, resolutions in the Exascale range are demanded since the simulation can easily entail six spatial decades (meters to microns).\\
Clearly, more sophisticated techniques, such as grid-refinement~\cite{ref:gridref}, or unstructured
grids, could significantly reduce the computational demand, but we shall not delve into these matters here.\\
Indeed, if the ``simple'' computational kernel is not performing well on today Petascale machines, there is no reason to expect it would scale up successfully to the Exascale. \\
Hence, we next discuss performance data for a Petascale class machine
(Marconi, Intel Phi based~\cite{ref:marconi}) for a VLB code.
Starting from the performance of a VLB, a rough estimate of more complex LB's can be provided:  a 
Shan-Chen multiphase flow \cite{ref:SHAN}, essentially two coupled VLB's, is about 2-3 times slower than VLB.\\
Our VLB code reached about 0.5 GLUPs for a single node, equivalent to about 50\% of the theoretical peak according to the roofline models\footnote{The nodes are Phi in cache mode, which means about 300 GB/s, hence an upper limit of about 1 GLUPS}.\\ 
In Fig.~\ref{fig:speedup1} the strong scaling is presented for a $ 512^3 $ resolution,
One point must be underlined: super-linear performance is due to the MCDRAM Memory limitation.
Below four nodes, memory falls short of storing the full problem, hence a slower standard RAM is instead used.
The best performance per node ($ \simeq 500 GLUPs $) is reached for a size of about $256^3$ per node, using $4$ tasks and $17$ threads per node.\\
In Fig.~\ref{fig:scaleup3} weak scaling results are presented using $ 128^3 $ gridpoints per task, 4 task per node, 17 threads per task.  A reasonably good performance, a parallel efficiency of $ 84\%$ is obtained using up to $700$ nodes, that means a 1.4 PFlops peak performance.\\
Three different levels of \textit{parallelism} have to be deployed for an efficient Exascale simulation:  
\begin{itemize}
\item At core/floating point unit level: (e.g., using vectorization)
\item At node level: (i.e., shared memory parallelization with threads)
\item At clustRoofKNL.jpeger level: (i.e., distributed memory parallelization with task) 
\end{itemize}
The user must be able to manage all the three levels but the third one
depends not only by the user (Sec.~\ref{Hiding}), since the
performance crucially depends on the technology and the topology of
the network, as well as on the efficiency of the communication software.

\section{Tentative ideas} \label{Future}
It is often argued that Exascale computing requires a new computational ``ecosystem'', meaning
by this that hardware, software and computational algorithms, must all
work in orchestrated harmony.

The challenges associated with achieving the above task on exascale machines is 
likely to require an optimal balance between evolutionary versus revolutionary changes to the 
existing algorithms, as well as  to the programming models. 
For very informative discussions on these topics, see \cite{DACOSTA,CAPGEIST,SNIR}.
Besides scalability, the authors point out a number of additional crucial metrics, such
as resilience to failure rates and silent errors, energy-consumption, runtime
reconfigurability, as well as new programming models and languages
aimed at supporting the above metrics.  
In the following, we present some potentially new ideas on the specific 
topics of communication/computation overlap and fault-tolerance in prospective
exascale LB applications. \\

\subsection{Time-delayed LB}
The idea of time-delayed PDE's as a means of further hiding communication costs
has been recently put forward by Ansumali and collaborators~\cite{ref:A2}.
The main point is to apply the Liouvillean operator $L$ 
at a delayed past time $t-\tau$, namely:
\begin{equation} 
\partial_t f(t) = L f(t-\tau)
\end{equation}
A formal Taylor expansion shows that this is tantamount to advancing the 
system at time $t$ via a retarded Liouvillean $L_{\tau} = L e^{-\tau L}$.\\
The delay provides a longer lapse between the posting of data request
(non-blocking send/receive) and the actual reception, thereby leaving more time for 
doing additional work, while waiting for the requested data.
In the case of LB, a M-level delayed streaming would look like follows
\begin{equation}
f_i(\vec{x},t+1) = \sum_{m=0}^M w_m f_i(\vec{x}-m\vec{c}_i,t-m)
\end{equation}
where $w_m$ are suitable normalized weights.\\
Incidentally, we note that for a negative-definite $L$, as required by stability,  
$L_{\tau}$ is a dilatation of $L$ for any $\tau>0$. 
Therefore, besides introducing extra higher-order errors, the delayed scheme
is subject to more stringent stability constraints than the standard one-time level version. \\
Nevertheless, once these issues are properly in place, delayed versions of LB might
prove very beneficial for Exascale implementations. 

\subsection{Fault-tolerant LB}
Another crucial issue in the Exascale agenda is to equip the computational model with 
error-recovery procedures, assisting the standard hardware operation 
against fault occurrences and silent errors (so far, hardware took up the full job).\\

Due to the large number of nodes/core, HW faults are very likely to appear routinely
by a mere statistical argument \cite{CAPGEIST}.
As a consequence, not only at system level all libraries \& API (e.g., MPI) have 
to be robust and fault-tolerant, but also the computational scheme.\\
To the best of our knowledge, for the case of LB, this key issue is entirely 
unexplored at the time of this writing.\\

\subsubsection{Check-Point Restart protocols}

The standard strategy to recover from system failures is Check-Point Restart (CPR), i.e.
roll the system back to the closest checkpoint and resume execution from there.
The CPR strategy is error-free, but very costly, since it requires a periodic dump
of the full system configuration, easily in the order of tens of thousands trillions variables
for Exascale applications (say, a cube of linear size $10^5$ lattice units).

The frequency at which CPR must take place is strictly related to the expected
failure rate of the system, which is a pretty complex quantity to assess.
However, at a failure rate of about two per hour,  as recently predicted on 
for exascale platforms\cite{SNIR}, an exaflops/s LB simulation with 
$10^{15}$ lattice sites, each taking $10^3$ flops/site/step, would advance
one timestep per second, which means  about $1$ CPR every $2000$ steps.
 Although this is well separated from the simulation timescale, it places nonetheless
a major strain on the I/O subsystem, as we are talking of the order of 1 Exabyte for each CPR.\\

A less-conservative but more sustainable procedure, consists of injecting a correction term at the time
when the error is first detected and keep going from there, without any roll-back, a
strategy sometimes  referred to as {\it imprecise computing}.\\
The close-eyes hope is that a suitable correction term can be extrapolated from the
past, and that the error still left shows capable of self-healing within a reasonably short transient time.\\
The first question is: how should such correction term look like?\\

Here comes a tentative idea.\\
Let $t_{CP}$ be the time of most recent check-point and $t_{ED}$ the time
of error-detection (ED). One could estimate the reconstructed ED state by 
time-extrapolating from $f(t_{CP})$ to $f(t_{ED})$.
Formally
\begin{equation} 
f^*(t_{ED}) = \mathcal{P} f(t_{CP}, t_1 \dots t_M)
\end{equation} 
where $\mathcal{P}$ denotes some suitable Predictor operator, extrapolating from $t_{CP}$ to $t_{ED}$.\\
This assumes the availability of a sequence of outputs  $f(t_m)$, where $t_1<t_2 \dots t_M < t_{CP}$.
Generally, that sequence is available but restricted to a set of hydrodynamic fields, say $h$ for short, such
as density and fluid velocity, namely four fields instead of twenty or so, major savings.
Density and velocity are sufficient to reconstruct the equilibrium component $f^{eq}$ but
not the full distribution $f=f^{eq} + f^{neq}$, because the non-equilibrium component depends 
on the gradients (in principle at all orders) of the equilibrium. 
One could then partially reconstruct $f^{neq}$ by taking first order gradients of density 
and velocity, a procedure which is going particularly touchy near the boundaries.
Here a peculiar feature of LB may prove pretty helpful: the first order gradients do not need to
be computed since they are contained in the momentum flux tensor, defined as
$ {\bf \Pi} = \sum_i f_i \vec{c}_i \vec{c}_i$, a {\it local} combination of the populations.\\
If the momentum flux tensor is also stored in the hydrodynamic set $ h= \lbrace \rho, \vec{u},{\bf \Pi} \rbrace$, then
the first-order gradient contribution to non-equilibrium is already accounted for, and only higher order 
non equilibrium terms, usually called ``ghosts'' in LB jargon, are left out from the reconstruction.
This is very good news; yet, it does not come for free, since the momentum flux tensor involves
another six fields, for a total of ten, only half of the memory required by the full $f$ configuration.
\\ 
Regardless of which policy is used, the entire procedure hangs on the 
(reasonable?) hope that leaving ghosts out of the reconstruction of $f(t_m)$,
is not going to compromise the quality of the predicted state $f^*(t_{ED})$.\\
Although only direct experimentation can tell, one may argue that numerical methods equipped 
with inherent stability properties, such as the entropic LB~\cite{ref:Entropic}, 
should stand good chances for fault-tolerant implementations.
For instance, restricting the predicted solution to feature an entropy increase (or 
non-decrease) as compared to the current state at time $t$, may help improve 
the quality of the reconstruction procedure. \\
It should also be pointed out that higher-order moments can also be systematically
reconstructed by suitable recursive procedures exploiting the recurrence 
relations between Hermite basis functions \cite{ref:RECURSION}.

\subsubsection{Mitigating Silent Data Corruptions}

The above procedures may prove pretty useful not only to design better CPR protocols, but also 
to mitigate the effects of silent errors, or better said, silent data corruptions (SDC's).
SDC's are expected to become a major source of concern on exascale systems for a
series of reasons, primarily the reduction in size and corresponding increase in number
of elementary components, which enhances their exposure to, say, cosmic radiation. 
A number of studies show that although the majority of SDC's leads to explicit crashes, 
a smaller fraction corrupts the results without crashing the application, which is of course
way more dangerous \cite{SNIR}.
To the best of our knowledge, the vulnerability of LB to SDC's stands completely 
unexplored at the time of this writing.
Here again, entropic formulations and/or recursive reconstruction of kinetic moments may offer new possibilities.
For instance, some errors might remain silent. i.e. escape detection, at the level of 
lowest order moments but not at the ghost level, so that ghosts could be used as SDC's detectors.  

Summarizing, the kinetic LB representation may offer new physics-inspired data-compression 
and error-detection/correction strategies to deal with fault-tolerance and silent data corruptions: 
the exploration of these strategies makes a very interesting 
research topic for the design of a new generation Exascale LB schemes.

\section{Prospective Exascale LB applications} \label{Summary}
\begin{figure}[h] \label{fig:COPMAT}
\includegraphics[scale=0.8]{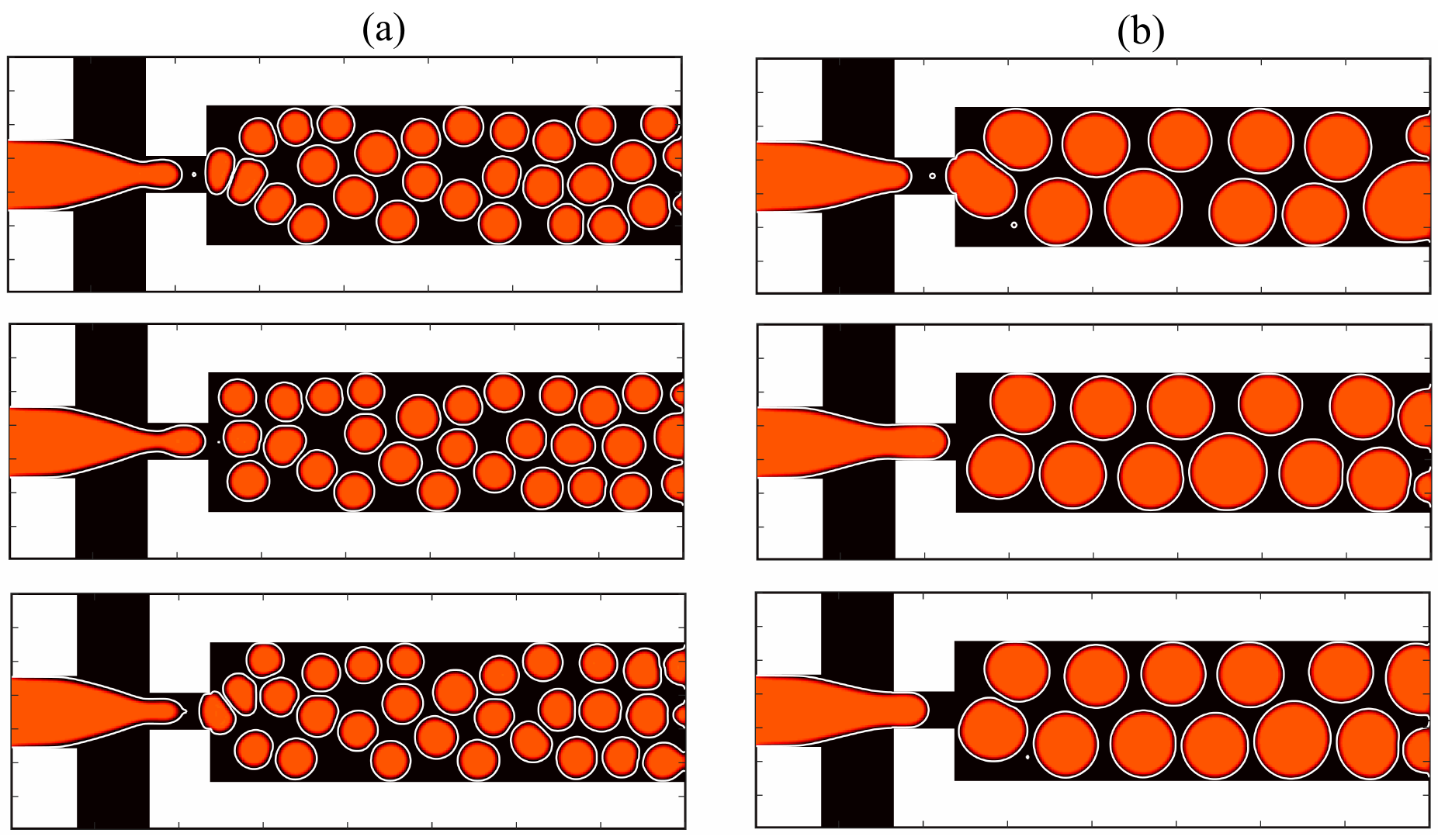}
\caption{Different configurations of emulsions. In panel (a) the continuous to dispersed flow rate ratio $Q^{in}_{d}/(2 Q^{in}_{c})=2$ whereas in panel (b) is set to one. The viscosities (in lattice unit) are the same for the two fluids and set to $\nu=0.0167$ and the velocities of the two fluids at the inlet channels are (in lattice units/step) (a): $u^{in}_{d}=0.01$, $u^{in}_{c}=0.01$, (b): $u^{in}_{d}=0.01$, $u^{in}_{c}=0.005$.}
\end{figure}

In the present paper, we have discussed at length the challenges faced by LB to meet
with Exascale requirements.
Given that, as we hope we have been able to show, this is not going to be
a plain ``analytic continuation of current LB's'', the obvious question before
concluding is: what can we do once Exascale LB is with us?
The applications are indeed many and most exciting.\\
For instance, Exascale LB codes will enable direct numerical simulations
at Reynolds numbers around $10^5 \div 10^6$, thereby relieving much work from turbulence models.
The full-scale design of automobiles will become possible.\\
At moderate Reynolds regimes, Exascale hemodynamics will permit to simulate a full heartbeat
at red-blood cell resolution in about half an hour, thus disclosing new
opportunities for precision medicine.\\
Likewise, Exascale LB-particle codes will allow millisecond simulations of protein dynamics within the 
cell, thereby unravelling invaluable information on many-body hydrodynamic effects under crowded conditions, 
which are vital to many essential biological functions, hence potentially also for medical therapies 
against neurological diseases \cite{ref:RMP2018}.\\
Further applications can be envisaged in micro-nanofluidics, such as the simulation of foams
and emulsions at near-molecular resolution, so as to provide a much more detailed
account of near-contact interface interactions, which are essential to understand the complex rheology of such flows.
For instance, the macroscale structure (say mm) of soft-flowing crystals, i.e., ordered collections of liquid droplets
within a liquid solvent (say oil in water), is highly sensitive to the nanoscale interactions which take place 
as interfaces come in near-contact, thereby configuring a highly challenging multiscale problem spanning
six spatial decades and nearly twice as many in time \cite{COPMAT}.\\ 
As a very short example, Fig.\ref{fig:COPMAT} shows different configurations of monodispersed emulsions at the outlet of 
the flow-focuser obtained by changing the dispersed/continuous flow rate ratios.\\
Such multiscale problem is currently handled via a combination of coarse-graining and grid-refinement techniques \cite{ref:AM2018}. 
Exascale computing would drastically relax the need for both, if not lift it altogether, thereby paving the way to
the computational design of mesoscale materials (mm) at near-molecular resolution (nm) \cite{ref:COPMAT}.\\
The prospects look bright and exciting, but a new generation of ideas and 
LB codes are needed to turn this mind-boggling potential into a solid reality.

\newpage


{}

\end{document}